\documentclass[aps,pra,preprint,groupedaddress,showpacs,amsmath,amssymb]{revtex4-1}
\usepackage{graphicx}
\usepackage{subfig}
\usepackage{amsmath}
\begin{document}

%\preprint{}

\title{Exceptional spectra of the two-qubit quantum Rabi model}


\author{Zhanyuan Yan}
\email{yanzhanyuan@ncepu.edu.cn}
%\thanks{}
\author{Bingbing Xu}
\email{479323309@qq.com}
\author{Zehui Yan}
%\thanks{}
\affiliation{School of Mathematics and Physics, North China Electrical Power University, Baoding 071000, China}

\date{\today}

\begin{abstract}
  In this report, we have studied exceptional spectra of the two-qubit quantum Rabi model in two situations. Firstly, an exceptional spectra is achieved in resonant condition, in which the frequencies of two qubit and photon field satisfy resonant relation. With transformed rotating-wave approximation (TRWA) method, the Rabi model can be mapped into the solvable formation in qubit-photon¡¯s Fock states space. Based on the quantum supperconducting circuits experiment setup, the best value range of the system parameters are discussed. In the "resonant station" working window, the energy spectrums are calculated. Secondly, a special quasi-exact solution of two qubits Rabi model in reservoir is achieved. The algebraic structure of Hamiltonian is analyzed in the photon number space, a closed quasi-exact eigenstates space is found, and the quasi-exact solution can be clearly found from the algebraic structure of Hamiltonian. The results could be used to test quantum Rabi model.
\end{abstract}

%\keywords{}

\maketitle

\section{INTRODUCTION}
The Rabi model describes the interaction between light and matter \cite{Rabi1936,Rabi1937}, which was introduced 70 years ago, it is one of the simplest and most universal models in modern physics. However, in quite a long period of time, people usually use J-C model\cite{J-C1963}, which can be achieved from Rabi model after taking the rotating-wave approximation(RWA), to describes the interaction between a two-level system and a quantized harmonic oscillator in the field of quantum optics, because it is hard to find a second conserved quantity besides the energy, and the Hamiltonian is often an infinite dimension Non-Diagonal matrix in Hilbert space. Recently, the interest on the quantum Rabi model has been relighted due to the realization of ultrastrong qubit-photon coupling in circuit QED experiments \cite{Niemczyk2010}, where the rotating-wave approximation breaks down, and the system¡¯s dynamics must be governed by the Rabi model. Numerical method is often used to obtain the energy spectrums and the wave functions \cite{Chen2008}. Till 2011, D. Braak used the property of $\mathbb{Z}_{2}$  symmetry of the Rabi model, an analytical solution was obtained using the Bargmann space of entire functions to model the bosonic degree of freedom \cite{Braak2011}. Furthermore, coherent states methods \cite{Chen2014}, perturbation method \cite{Yu2012}, Bogoliubov operators \cite{Peng2015,Chen2012}. However, the problem is far from solved, the analytical solution is often in the form of series expansion, and it is difficult to extract the fundamental physics from results of the Rabi model and to precisely control the experimental parameters to process quantum information \cite{You2011}.

Here, we are interested in the quantum Rabi model for two qubits, it is basic and fundamental to the construction of the universal quantum gate \cite{Barends2014}. Recently, theoretical prediction and experimental realization of new phenomena or special characters related to two-qubit Rabi model are extensively concerned. In experiment, many two-qubit gates for superconducting qubits require specific arrangements of qubit frequencies to perform optimally \cite{Sheldon2016}. For example, The cross resonance (CR) gate works for qubits within a narrow window of detunings defined by the anharmonicity of the qubits, this restriction becomes accentuated in larger networks of qubits where all qubit frequencies must be arranged within a small frequency window \cite{Chow2011}. Its analytical solution was obtained in \cite{Peng2014} by means of Bargmann space approach, the exceptional solutions in \cite{Chilingaryan2015}, and also in \cite{Chen2014} with extended coherent states representation. However, the analytical solutions expressed in qubit-photon¡¯s Fock states space are expected, because it is more fundamental in physics. With a perturbation theory, L Yu has achieved the analytical solutions of Rabi model that describes the interaction between a two-level system and a quantized harmonic oscillator in photon number space \cite{Yu2012}. In this report, we extend this method to two-qubit circumstance. With transformed rotating-wave approximation (TRWA), the analytical solutions for two qubits Rabi model are achieved in resonant condition. Besides, a special ¡°dark states¡±-like eigenstate of two-qubit quantum Rabi model is found in \cite{Peng2014}, which may provide some interesting application in a simpler way. In this paper, we concern about the exceptional spectra of two-qubit Rabi model in a reservoir, and the characters of quasi-exact solutions and algebraic structure of Hamiltonian.

The paper is organized as follows. In section II, we analytically retrieve the solution of the two-qubit quantum Rabi model using TRWA method. In section III, the ¡°dark states¡±-like eigenstate of two-qubit Rabi model in reservoir is studied. Finally, we make some conclusions in section IV.

\section{SOLUTION OF TWO-QUBIT QUANTUM RABI MODEL IN QUBIT-PHOTON'S FOCK STATES SPACE}
Now, we will consider the two-qubit Rabi model which has the Hamiltonian of the form
\begin{equation}
H = \omega {a^ {\dag} }a + {g_1}{\sigma _{1x}}(a + {a^ {\dag} }) + {g_2}{\sigma _{2x}}(a + {a^ {\dag} }) + {\Delta _1}{\sigma _{1z}} + {\Delta _2}{\sigma _{2z}},
\end{equation}
where
$\sigma _{1x}$, $\sigma _{2x}$, $\sigma _{1z}$, $\sigma _{2z}$  are the Pauli matrices for the two two-level atoms with level splitting
$\Delta_1$, $\Delta_2$ , respectively, $g_1$ and $g_2$ are the coupling parameters.
When performing a rotation around the $y$ axis with $R = {e^{i\pi \frac{{{\sigma _{1y}}}}{4}}} \otimes {e^{i\pi \frac{{{\sigma _{2y}}}}{4}}}$, the Rabi model becomes
\begin{equation}
{H'} = \omega {a^{\dag} }a - {g_1}{\sigma _{1z}}(a + {a^ {\dag} }) - {g_2}{\sigma _{2z}}(a + {a^ {\dag} }) + {\Delta _1}{\sigma _{1x}} + {\Delta _2}{\sigma _{2x}}.
\end{equation}
Based on a unitary transformation, the transformed rotating-wave approximation (TRWA) method is proposed to study spin-boson system \cite{Zhao2011}, this approach takes into account the effect of counter-rotating terms but still keeps the Hamiltonian with a simple mathematical structure. The unitary transformation read as
\begin{equation}
{U_1} = {e^{{\lambda _1}{\sigma _{1z}}({a^{\dag} } - a)}}, {U_2} = {e^{{\lambda _2}{\sigma _{2z}}({a^{\dag} } - a)}},
\end{equation}
with $\lambda_1$ and $\lambda_2$ being the dimensionless parameter determined by the following calculations, an effective Hamiltonian is given by ${H_E} = U_2^ {\dag} U_1^{\dag} {H'}{U_1}{U_2}$, namely,
\begin{equation}
{H_E} = {H_1} + {H_2} + {H_3},
\end{equation}
where
\begin{subequations}
\begin{eqnarray}
{H_1}=&& \omega {a^ {\dag} }a - {\lambda _2}\omega {\sigma _{2z}}({a^ {\dag} } + a) + \lambda _{^2}^2\omega  - {\lambda _1}\omega {\sigma _{1z}}({a^{\dag} } + a) + 2{\lambda _1}{\lambda _2}\omega {\sigma _{2z}}{\sigma _{1z}}{\rm{ + }}\lambda _1^2\omega,\\
{H_2} =&& - {g_1}({a^ {\dag} } + a){\sigma _{1z}} + 2{\lambda _2}{g_1}{\sigma _{1z}}{\sigma _{2z}} + 2{\lambda _1}{g_1} - {g_2}({a^ {\dag} } + a){\sigma _{2z}} + 2{\lambda _2}{g_2},\\
{H_3}=&&{\Delta _1}\left\{ {{\sigma _{1x}}consh\left[ {2{\lambda _1}({a^ {\dag} } - a)} \right] + i{\sigma _{1y}}\sinh \left[ {2{\lambda _1}({a^ {\dag} } - a)} \right]} \right\}\nonumber\\
&&+ {\Delta _2}\left\{ {{\sigma _{2x}}consh\left[ {2{\lambda _2}({a^ {\dag} } - a)} \right] + i{\sigma _{2y}}\sinh \left[ {2{\lambda _2}({a^ {\dag} } - a)} \right]} \right\}.
\end{eqnarray}
\end{subequations}
In Eq.(5C), if the multiphoton process is neglected, then the high-order terms for $a$ and ${a^\dag}$ can be eliminated. In the eigenstates of ${\sigma _{1x}}$ with ${\sigma _{1x}}\left|  \pm  \right\rangle {\rm{ = }} \pm \left|  \pm  \right\rangle $  and in the eigenstates of ${\sigma _{2x}}$ with ${\sigma _{2x}}\left|  \pm  \right\rangle {\rm{ = }} \pm \left|  \pm  \right\rangle $, see appendix A, the effective Hamiltonian Eq.(4) reduces to the form
\begin{eqnarray}
H_{E} =&& \omega {a^{\dag} }a + \lambda _1^2\omega  + 2{\lambda _1}{g_1} + 2{\lambda _2}{g_2} \nonumber\\
&&+ \lambda _{^2}^2\omega  + (2{\lambda _2}{g_1} + 2{\lambda _1}{g_2} + 2{\lambda _1}{\lambda _2}\omega )({\tau _{1 + }} + {\tau _{1 - }})({\tau _{2 + }} + {\tau _{2 - }})\nonumber\\
 &&+ ({g_1} + {\lambda _1}\omega )({\tau _{1 + }} + {\tau _{1 - }})({a^ {\dag} } + a) + ({g_2} + {\lambda _2}\omega )({\tau _{2 + }} + {\tau _{2 - }})({a^{\dag} } + a)\nonumber\\
 &&+ {\Delta _1}\left\{ {{\tau _{1z}}{G_0}(N) + ({\tau _ + } - {\tau _ - }){\rm{[}}{{\rm{F}}_1}(N){a^ {\dag} } - a{F_1}(N)]} \right\}\nonumber\\
 &&+ {\Delta _2}\left\{ {{\tau _{2z}}{G_0}^\prime (N) + ({\tau _ + } - {\tau _ - }){\rm{[}}{{\rm{F}}_1}^\prime (N){a^ {\dag} } - a{F_1}^\prime (N)]} \right\}.
\label{Heff}
\end{eqnarray}
The effective Hamiltonian conserves $\mathbb{Z}_{2}$ symmetry with the transformation $R'={e^{i\pi {a^ {\dag} }a}}\otimes \sigma_{1x}\otimes \sigma_{2x}$, which allows us to extend the idea of a parity basis used in the single-qubit Rabi model\cite{Casanova2010}. In two qubits system, Hilbert space also split in two unconnected subspaces or parity chains as
\begin{subequations}
\begin{eqnarray}
{\left| j \right\rangle _ + } = \left\{ \begin{array}{l}
 \cdots \left| {2n + 1, - , + } \right\rangle ,\left| {2n + 1, + , - } \right\rangle ,\left| {2n + 2, + , + } \right\rangle ,\\
{\kern 1pt} {\kern 1pt} {\kern 1pt} {\kern 1pt} {\kern 1pt} {\kern 1pt} {\kern 1pt} {\kern 1pt} {\kern 1pt} {\kern 1pt} {\kern 1pt} \left| {2n + 2, - , - } \right\rangle ,\left| {2n + 3, - , + } \right\rangle ,\left| {2n + 3, + , - } \right\rangle  \cdots
\end{array} \right\},\\
{\left| j \right\rangle _ - } = \left\{ \begin{array}{l}
 \cdots \left| {2n, - , + } \right\rangle ,\left| {2n, + , - } \right\rangle ,\left| {2n + 1, + , + } \right\rangle ,\\
{\kern 1pt} {\kern 1pt} {\kern 1pt} {\kern 1pt} {\kern 1pt} {\kern 1pt} {\kern 1pt} {\kern 1pt} {\kern 1pt} {\kern 1pt} {\kern 1pt} \left| {2n + 1, - , - } \right\rangle ,\left| {2n + 2, - , + } \right\rangle ,\left| {2n + 2, + , - } \right\rangle  \cdots
\end{array} \right\}.
\end{eqnarray}
\end{subequations}
In Fock space, the Hamiltonian is infinite dimensional with off-diagonal elements, see appendix A. In general, it cannot be solved analytically. However, in the current experimental setup with ultrastrong coupling ($\lambda  < 0.5\omega$), the numerical result shows that the dimensionless parameter $\lambda$ is small compared with the unit. In single qubit Rabi model, if $({g} + {\lambda}\omega )\sqrt {2n + 1}  - {\Delta}{F}(2n + 1,2n) = 0$, the Rabi model can be mapped into the solvable Jaynes-Cummings-like model, in Ref.\cite{Yu2012}, the dimensionless parameter $\lambda$ is chosen as
${\lambda} \approx  - \frac{{{g}}}{{\omega  + 2{\Delta}{e^{ - {{(\frac{g}{{\omega  + 2{\Delta }}})}^2}}}}}$, the ground and excited-state-energy spectrums agree well with the direct numerical simulation in a wide range of the experimental parameters. For two-qubit Rabi model, choosing proper parameter $\lambda_1$ and $\lambda_2$, let
\begin{equation}
({g_1} + {\lambda _1}\omega ) + 2{\Delta _1}{\lambda _1}{e^{ - 2\lambda _1^2}}{\rm{ = }}0,
\label{lambda1}
\end{equation}
\begin{equation}
({g_2} + {\lambda _2}\omega ) - 2{\Delta _2}{\lambda _2}{e^{ - 2\lambda _2^2}}{\rm{ = }}0.
\label{lambda2}
\end{equation}
the structure of Hamiltonian (A2) will become much more simple, but cannot be solved analytically yet. However, if the frequency $\omega$, $\Delta _1$ and $\Delta _2$ are proper controlled in experiment, and satisfied the relation
\begin{equation}
2{\lambda _2}{g_1} + 2{g_2}{\lambda _1} + 2{\lambda _1}{\lambda _2}\omega  = 0.
\label{resonant}
\end{equation}
the Hamiltonian matrix takes the formation of block diagonal matrix, each block is a $4\times 4$ matrix, read as
\begin{equation}
_ + \left\langle j \right|{H_E}{\left| j \right\rangle _ + } = \left( {\begin{array}{*{20}{c}}
 \ddots & \vdots & \vdots & \vdots & \vdots & \vdots & \vdots & {\mathinner{\mkern2mu\raise1pt\hbox{.}\mkern2mu
 \raise4pt\hbox{.}\mkern2mu\raise7pt\hbox{.}\mkern1mu}} \\
 \cdots &A&0&0&0&0&0& \cdots \\
 \cdots &0&B&X&Y&0&0& \cdots \\
 \cdots &0&X&C&0&Y&0& \cdots \\
 \cdots &0&Y&0&D&X&0& \cdots \\
 \cdots &0&0&Y&X&{A'}&0& \cdots \\
 \cdots &0&0&0&0&0&{B'}& \cdots \\
 {\mathinner{\mkern2mu\raise1pt\hbox{.}\mkern2mu
 \raise4pt\hbox{.}\mkern2mu\raise7pt\hbox{.}\mkern1mu}} & \vdots & \vdots & \vdots & \vdots & \vdots & \vdots & \ddots
\end{array}} \right)
\label{block diagonal Matrix}
\end{equation}
in which
\begin{eqnarray*}
X =&& ({g_2} + {\lambda _2}\omega )\sqrt {2n + 2}  + {\Delta _2}{F_1}^\prime (2n + 2,2n + 1),\\
Y = &&({g_1} + {\lambda _1}\omega )\sqrt {2n + 2}  - {\Delta _1}{F_1}(2n + 2,2n + 1),
\end{eqnarray*}
\begin{eqnarray*}
A = &&\omega (2n + 1) + \lambda _1^2\omega  + \lambda _{^2}^2\omega  + 2{\lambda _1}{g_1} + 2{\lambda _2}{g_2} - {\Delta _1}{G_0}(2n + 1) + {\Delta _2}{G_0}(2n + 1),\\
B =&& (2n + 1)\omega  + \lambda _1^2\omega  + \lambda _{^2}^2\omega  + 2{\lambda _1}{g_1} + 2{\lambda _2}{g_2} + {\Delta _1}{G_0}(2n + 1) - {\Delta _2}{G_0}^\prime (2n + 1),
\end{eqnarray*}
\begin{eqnarray*}
C =&& (2n + 2)\omega  + \lambda _1^2\omega  + \lambda _{^2}^2\omega  + 2{\lambda _1}{g_1} + 2{\lambda _2}{g_2} + {\Delta _1}{G_0}(2n + 2) + {\Delta _2}{G_0}^\prime (2n + 2),\\
D = &&(2n + 2)\omega  + \lambda _1^2\omega  + \lambda _{^2}^2\omega  + 2{\lambda _1}{g_1} + 2{\lambda _2}{g_2} - {\Delta _1}{G_0}(2n + 2) - {\Delta _2}{G_0}^\prime (2n + 2),
\end{eqnarray*}
\begin{eqnarray*}
A' =&& \omega (2n + 3) + \lambda _1^2\omega  + \lambda _{^2}^2\omega  + 2{\lambda _1}{g_1} + 2{\lambda _2}{g_2} - {\Delta _1}{G_0}(2n + 3) + {\Delta _2}{G_0}(2n + 3),\\
B' =&& \omega (2n + 3) + \lambda _1^2\omega  + \lambda _{^2}^2\omega  + 2{\lambda _1}{g_1} + 2{\lambda _2}{g_2} + {\Delta _1}{G_0}(2n + 3) - {\Delta _2}{G_0}^\prime (2n + 3).
\end{eqnarray*}
The wave function, correspondence to the block matrix, is
\begin{eqnarray}
\left| \psi  \right\rangle  =\dots &&+ {E_{2n + 1, + , - }}\left| {2n + 1, + , - } \right\rangle  + {E_{2n + 2, + , +}}\left| {2n + 2, + , + } \right\rangle  \nonumber\\
&&+ {E_{2n + 2, - , - }}\left| {2n + 2, - , - } \right\rangle  + {E_{2n + 3, - , + }}\left| {2n + 3, - , + } \right\rangle \dots.
\end{eqnarray}
Then, the Hamiltonian matrix of two qubits Rabi model in Fock space cannot be solved analytically.

Eq.(\ref{lambda1}), (\ref{lambda2}), (\ref{resonant}) suggests a special condition, a kind of "resonant state" relies on experimental ability, the detailed value of parameter $g_1, g_2, \Delta_1, \Delta_2$ is the working window of the system, the energy spectrum of "resonant state" could be used to examine Rabi model.

For the superconducting circuits are currently the most experimentally advanced solid-state qubits, the following discussion about the parameter of the "resonant state" based on the experimental setup in ref.\cite{Niemczyk2010}, with ultrastrong coupling $0.1\le g_1,g_2\le 1.0$ . In the process of TRWA, $\lambda_1$ and $\lambda_1$ have been regarded as small parameter, so we make the absolute value of them less than 0.1 to ensure the accuracy of approximation. Furthermore, we assume that the second Josephson junction is designed firstly with $\Delta_2$ varied from $1.0\omega $ to $2.5\omega$, the other parameters are designed according to resonant relations Eq.(\ref{lambda1}), (\ref{lambda2}), (\ref{resonant}), and $g_1$ is the last adjustable variable.

From the structure of equation (\ref{lambda1}), (\ref{lambda2}), (\ref{resonant}), we find Eq.(\ref{lambda2}) is the most critical relation, for the existence of singularities in the solution of $\lambda_2$, which is not exist in Eq.(\ref{lambda1}). So, we begin with the discussion of appropriate value of $g_2$ determined by $\lambda_2$.
\begin{figure}[h]
	\centering
	\subfloat[]{\includegraphics[width=.45\columnwidth]{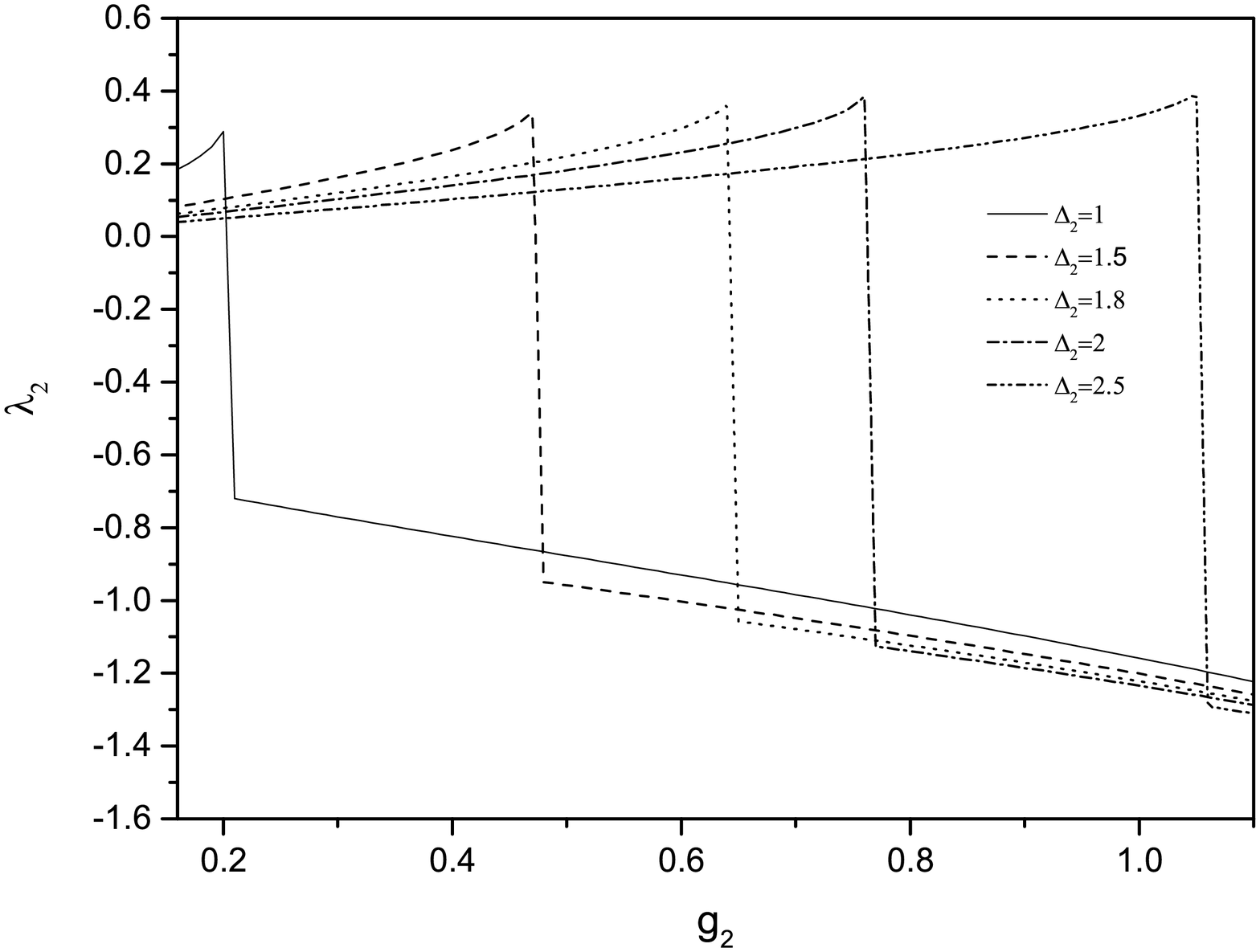}} \qquad
	\subfloat[]{\includegraphics[width=.45\columnwidth]{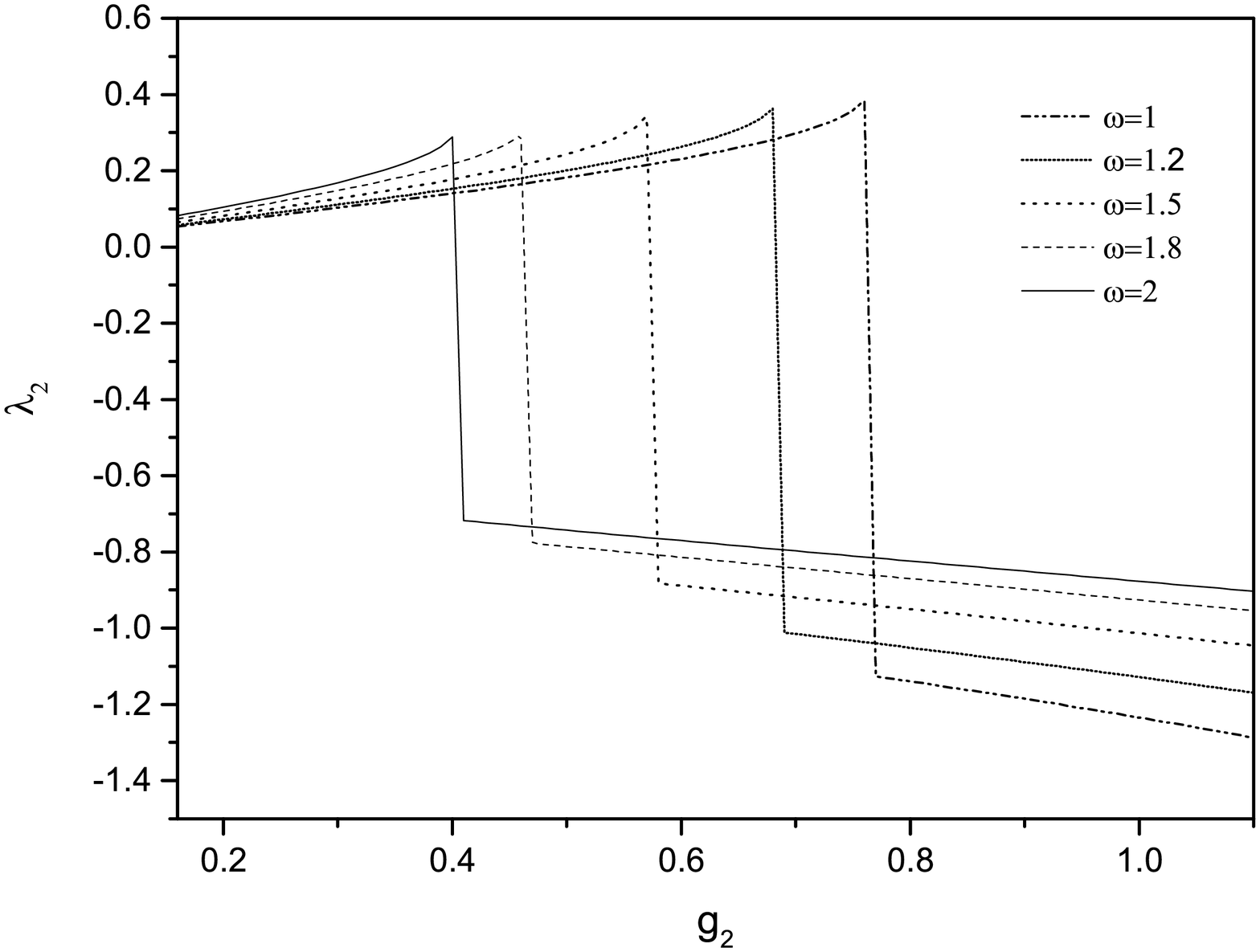}}
	\caption{(a)  The range of $\lambda_2$ with varied $\Delta_2$, when $\omega=1.0$. (b) The range of $\lambda_2$ with varied $\omega$, when $\Delta_2=2.0$.}
	\label{fig:g2lambda2}
\end{figure}

 A sharp decline of $\lambda_2$ is found in Fig.1(a), with $\Delta_2=1.0,1.5,2.0,2.5$. With the statement above, when the absolute value of $\lambda_2$ is larger than 0.1, the approximation is failure. Therefor, a larger value of $\Delta_2$ is benefit to the selecting of $g_2$, when $\Delta_2$ near to 2.5, the coupling parameter $g_2$ can be chosen between 0.1 to 1.0 in the case of $\omega=1.0$. if $\Delta_2$ is determined, the range of $g_2$ shrinked with the decrease of $\omega$ as shown in Fig.1(b), with the typical frequency $\omega=1.0$, the range of $g_2$ is [0.1,0.8]. The same situation also appears in the discussion of $\Delta_1$ indicate by Fig.2, because resonant relation Eq.(\ref{resonant}) transmits the character of $\lambda_2$ to $\Delta_1$. The value of $\Delta_1$ should be in the region [0.05,0.4] to satisfy the resonant relation, when $g_1=0.9$.
\begin{figure}[h]
	\centering
	\subfloat[]{\includegraphics[width=.45\columnwidth]{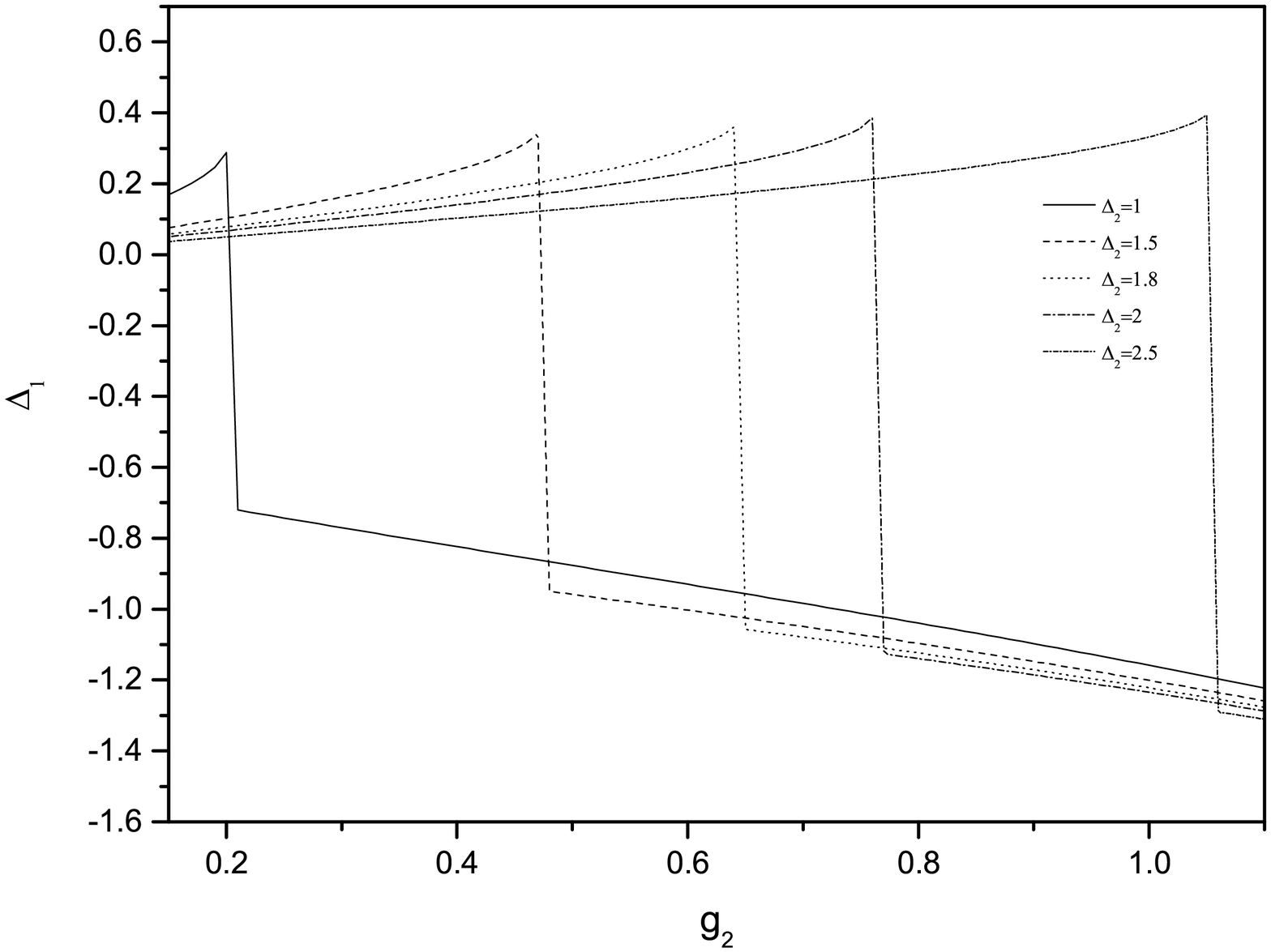}} \qquad
	\subfloat[]{\includegraphics[width=.45\columnwidth]{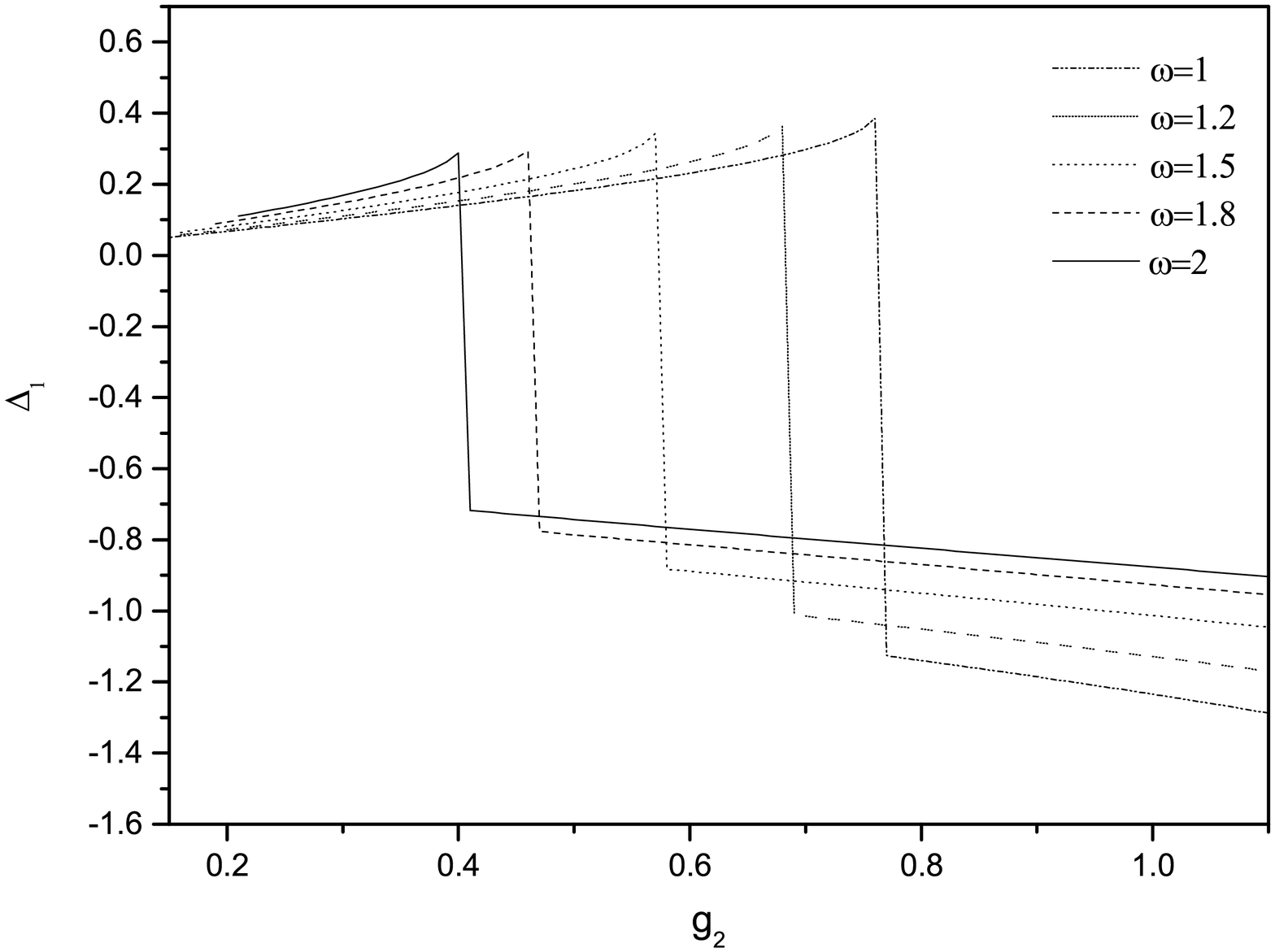}}
	\caption{(a)  The range of $\Delta_1$ with varied $\Delta_2$, when $\omega=1.0$,$g_1=0.9$. (b) The range of $\Delta_1$ with varied $\omega$, when $\Delta_2=2.0$,$g_1=0.9$.}
	\label{fig:g2delta1}
\end{figure}

In the "resonant state" working window of the system, the energy spectrum are plotted in Fig.3 with $\omega=1.0$ for a typical frequency, and $\Delta_2=2.0\omega$, $g_2=0.7$, to ensure the working window as large as possible. The energy spectrum could be checked experimentally, furthermore, two-qubit quantum Rabi model could be examined.
\begin{figure}[h]
\includegraphics[width=12cm]{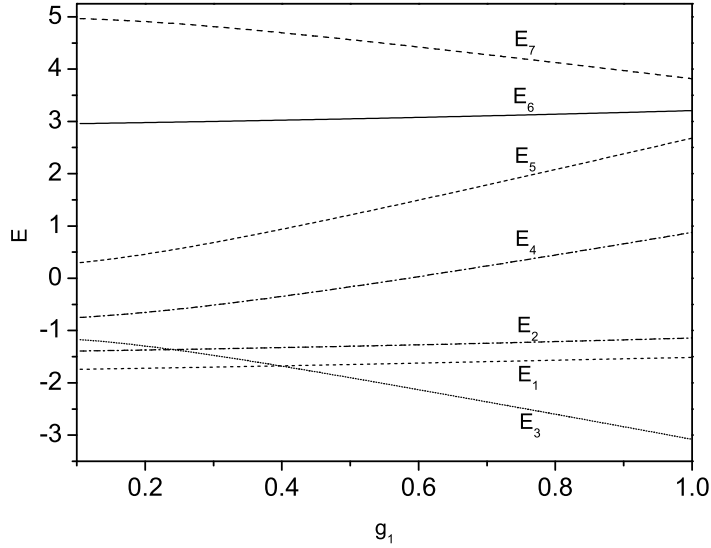}
\caption{\label{Fig:energy spectrum} "Resonant state" energy spectrum of two-qubit quantum Rabi model.}
\end{figure}

\section{QUASI-EXACT SOLUTION of TWO-QUBIT QUANTUM RABI MODEL IN RESERVOIR }
Realistic quantum systems cannot avoid interactions with their environments, thus the study of open quantum systems is very important. Many efforts have been focused on decoherence and disentanglement effects in Markovian process and non-Markovian processes \cite{Breuer2009,Rivas2010,Haikka2011,Jing2010,Chen2016}, However, the algebraic structure of Hamiltonian of the multi qubit system is seldom studied before. In this section, we will discuss the exceptional spectra of two-qubit quantum Rabi model in reservoir, the Hamiltonian of the system is given by
\begin{eqnarray}
{H_{st}} =&& \omega {a^ + }a + \sum\limits_k {{\omega _k}b_k^ + {b_k}}  + \sum\limits_k {{V_k}} (b_k^ + a + {a^ + }{b_k}) + {g_1}{\sigma _{1x}}(a + {a^ + }) + {g_2}{\sigma _{2x}}(a + {a^ + })\nonumber\\
&&  + {\Delta _1}{\sigma _{1z}} + {\Delta _2}{\sigma _{2z}} + \sum\limits_k {{g_{1k}}^\prime (} {b_k} + b_{_k}^ + ){\sigma _{1x}} + \sum\limits_k {{g_{2k}}^\prime (} {b_k} + b_{_k}^ + ){\sigma _{2x}}.
\end{eqnarray}
in which, ${\omega _k}$, $b_k$ and $b_k^{\dag}$ are respectively the frequency, annihilation and creation operators for the k-th harmonic oscillator of the reservoir, and $g'_{1k}$ and $g'_{2k}$ are the coupling parameters between the qubits and the environment. $V_k$ is the coupling constant between the k-th harmonic oscillator of the zero-temperature bosonic reservoir and the transmission line resonator.

With the pseudomode approach, the number of the pseudomodes relies on the shape of the reservoir spectral distribution. Since a Lorentzian function
\begin{equation}
D(\omega ) = \frac{\Gamma }{{\left( {\omega  - {\omega _c}} \right) + {{\left( {\frac{\Gamma }{2}} \right)}^2}}}.
\end{equation}
has only one pole in the lower half complex plane. According to the theory of the pseudomode, the number of the pseudomode depends on the form of the spectral distribution of the reservoir. For the single Lorentzian model, the interaction between the reservoir and the quantum system in the whole spectrum is represented by the interaction of a pseudomode at the singular point. The Hamiltonian describing the interaction between two qubits and the reservoir and the optical field is as follow:
\begin{eqnarray}
H' =&& \omega {a^ {\dag} }a + {\omega _1}{b^ {\dag} }b + V({b^ {\dag} }a + {a^ {\dag} }b) + {g_1}{\sigma _{1x}}(a + {a^ {\dag} }) + {g_2}{\sigma _{2x}}(a + {a^ {\dag} })\nonumber\\
 &&+ {\Delta _1}{\sigma _{1z}} + {\Delta _2}{\sigma _{2z}} + {g_1}^\prime {\sigma _{1x}}(b + {b^ {\dag} }) + {g_2}^\prime {\sigma _{2x}}(b + {b^ {\dag} }).
\end{eqnarray}
On the basis of the section II, an additional transformation $U' = {e^{\lambda_ 1 {\sigma _{1x}}({b^ {\dag} } - b)}}{e^{\lambda_ 2 {\sigma _{2x}}({b^ {\dag} } - b)}}$ is put on $H'$, the effective Hamiltonian can be written as
\begin{eqnarray}
{H'_{E}} =&& \omega {a^{\dag} }a + {\omega _1}{b^ {\dag} }b + \lambda _1^2\omega  + \lambda _{^2}^2\omega  + 2{\lambda _1}{g_1} + 2{\lambda _2}{g_2}\nonumber\\
 &&+ {g_1}^\prime (b + {b^ {\dag} })({\tau _{1 + }} + {\tau _{1 - }}) + {g_2}^\prime (b + {b^ {\dag} })({\tau _{2 + }} + {\tau _{2 - }})\nonumber\\
 &&+ ({g_2} + {\lambda _2}\omega )({\tau _{2 + }}{a^ {\dag} } + {\tau _{2 - }}a) + ({g_1} + {\lambda _1}\omega )({\tau _{1 + }}{a^ {\dag} } + {\tau _{1 - }}a)\nonumber\\
 &&+ {\Delta _1}\left\{ {{\tau _{1z}}{G_0}(N) + {\rm{[}} + {\tau _{1{\rm{ + }}}}{{\rm{F}}_1}(N){a^{\dag} } + {\tau _{1 - }}a{F_1}(N)]} \right\}\nonumber\\
 &&+ {\Delta _2}\left\{ {{\tau _{2z}}{G_0}^\prime (N) + {\rm{[}}{{\rm{F}}_1}^\prime (N){\tau _{2 + }}{a^ {\dag} }{\rm{ + }}{\tau _{2 - }}a{F_1}^\prime (N)]} \right\}
\end{eqnarray}
$H'_{E}$ process a $\mathbb {Z}_2$ symmetry with the transformation $R''={e^{i\pi {a^ {\dag} }a}}\otimes {e^{i\pi {b^ {\dag} }b}}\otimes \sigma_{1x}\otimes \sigma_{2x}$. Taking odd parity, ($m+n$ is even), Hilbert space for example,
\begin{eqnarray}
\left| J \right\rangle  = \left[ \begin{array}{l}
 \cdots \left| {m,n - 1, -,  - } \right\rangle ,\left| {m - 1,n, + , + } \right\rangle ,\left| {m,n, + , - } \right\rangle ,\left| {m,n, -,  + } \right\rangle ,\\
{\kern 1pt} {\kern 1pt} {\kern 1pt} {\kern 1pt} {\kern 1pt} {\kern 1pt} {\kern 1pt} {\kern 1pt} {\kern 1pt} {\kern 1pt} {\kern 1pt} \left| {m + 1,n, - , - } \right\rangle ,\left| {m,n + 1, + , + } \right\rangle ,\left| {m + 1,n + 1, + , - } \right\rangle ,\\
{\kern 1pt} {\kern 1pt} {\kern 1pt} {\kern 1pt} {\kern 1pt} {\kern 1pt} {\kern 1pt} {\kern 1pt} {\kern 1pt} {\kern 1pt} {\kern 1pt} \left| {m + 1,n + 1, - , + } \right\rangle ,\left| {m + 2,n + 1, - , - } \right\rangle ,\left| {m + 1,n + 2, + , + } \right\rangle  \cdots
\end{array} \right]
\end{eqnarray}

The effective Hamiltonian matrix takes the formation as
{\scriptsize{\begin{eqnarray}
\left( {\begin{array}{*{20}{c}}
&\vdots &{}&{}&{}&{}&{}&{}&{}&{}& \vdots \\
&A&0&{\sqrt n {K_1}}&{\sqrt n {K_2}}&{0}&{0}&{0}&{0}&{0}&{0}\\
&0&B&{{g_2}^\prime \sqrt m }&{{g_1}^\prime \sqrt m }&{0}&{0}&{0}&{0}&{0}&{0}\\
&{\sqrt n {K_1}}&{{g_2}^\prime \sqrt m }&C&0&{{g_1}^\prime \sqrt {m + 1} }&{\sqrt {n + 1} {K_2}}&{0}&{0}&{0}&{0}\\
&{\sqrt n {K_2}}&{{g_1}^\prime \sqrt m }&0&D&{{g_2}^\prime \sqrt {m + 1} }&{\sqrt {n + 1} {K_1}}&{0}&{0}&{0}&{0}\\
&{0}&{0}&{{g_1}^\prime \sqrt {m + 1} }&{{g_2}^\prime \sqrt {m + 1} }&E&0&{\sqrt {n + 1} {K_1}}&{\sqrt {n + 1} {K_2}}&{0}&{0}\\
&{0}&{0}&{\sqrt {n + 1} {K_2}}&{\sqrt {n + 1} {K_1}}&0&F&{\sqrt {m + 1} {g_2}}&{\sqrt {m + 1} {g_1}}&{0}&0{}\\
&{0}&{0}&{0}&{0}&{\sqrt {n + 1} {K_1}}&{\sqrt {m + 1} {g_2}}&G&0&{\sqrt {m + 2} {g_1}}&{\sqrt {n + 2} {K_2}}\\
&{0}&{0}&{0}&{0}&{\sqrt {n + 1} {K_2}}&{\sqrt {m + 1} {g_1}}&0&H&{\sqrt {m + 2} {g_2}}&{\sqrt {n + 2} {K_1}}\\
&{0}&{0}&{0}&{0}&{0}&{0}&{\sqrt {m + 2} {g_1}}&{\sqrt {m + 2} {g_2}}&I&0\\
&{0}&{0}&{0}&{0}&{0}&{0}&{\sqrt {n + 2} {K_2}}&{\sqrt {n + 2} {K_1}}&0&J\\
& \vdots &{}&{}&{}&{}&{}&{}&{}&{}& \vdots
\end{array}} \right)
\end{eqnarray}}}
in which,
\[{K_1} = {g_1} - \frac{{{g_1}}}{{\omega  - {\eta _1}}}\omega  - 2{\Delta _1}\frac{{{g_1}}}{{\omega  - {\eta _1}}}{e^{ - 2{{(\frac{{{g_1}}}{{\omega  - {\eta _1}}})}^2}}},\]
\[{K_2} = {g_2} - \frac{{{g_2}}}{{\omega  - {\eta _2}}}\omega  - 2{\Delta _2}\frac{{{g_2}}}{{\omega  - {\eta _2}}}{e^{ - 2{{(\frac{{{g_2}}}{{\omega  - {\eta _2}}})}^2}}}.\]
with ${\eta _i} = 2{\Delta _i}{e^{ - {{(\frac{{{g_i}}}{{\omega  - 2{\Delta _i}}})}^2}}}, (i = 1,2)$.

In this subspace, the initial state $\left| \psi  \right\rangle $ is expressed with the coefficient
\begin{eqnarray}
\left\{ \begin{array}{l}
 \cdots {c_{m + n - 1,1}},{c_{m + n - 1,2}},{c_{m + n,1}},{c_{m + n,2}},{c_{m + n + 1,1}},\\
{\kern 1pt} {\kern 1pt} {\kern 1pt} {\kern 1pt} {\kern 1pt} {\kern 1pt} {\kern 1pt} {\kern 1pt} {\kern 1pt}{\kern 1pt}{\kern 1pt}{\kern 1pt}{\kern 1pt} {\kern 1pt} {\kern 1pt} {c_{m + n + 1,2}},{c_{m + n + 2,1}},{c_{m + n + 2,2}},{c_{m + n + 3,1}},{c_{m + n + 3,2}} \cdots
\end{array} \right\}
\label{truncation}
\end{eqnarray}
For $H\left| \psi  \right\rangle  = \left| {\psi '} \right\rangle $, we study the quasi-exact eigenstates\cite{Peng2015}, with the coefficients of ${g_1}^\prime  = {g_2}^\prime ,{g_1}{\rm{ = }}{g_2},{\Delta _1} = {\Delta _2}$, furthermore,
\begin{eqnarray}
{g_1}^\prime \sqrt {m + 1} {c_{m + n,1}} + {g_2}^\prime \sqrt {m + 1} {c_{m + n,2}} = 0,\nonumber\\
\sqrt {n + 1} {K_2}{c_{m + n,1}} + \sqrt {n + 1} {K_1}{c_{m + n,2}} = 0,
\end{eqnarray}
then this subspace is closed. By using the time independent Sch\"{o}dinger equation, we obtain
\begin{eqnarray}
\sqrt n {K_1}{c_{m + n - 1,1}} + {g_2}^\prime \sqrt m {c_{m + n - 1,2}} + ({\omega _1}m + \omega n){c_{m + n,1}} = E{c_{m + n,1}},\nonumber\\
\sqrt n {K_2}{c_{m + n - 1,1}} + {g_1}^\prime \sqrt m {c_{m + n - 1,2}} + ({\omega _1}m + \omega n){c_{m + n,2}} = E{c_{m + n,2}}.
\end{eqnarray}
So, when the eigen energy $E = {\omega _1}m + \omega n$, for a special case, ${c_{m + n - 1,1}} = {c_{m + n - 1,2}} = 0$, there is a invariant subspace formed
by $\left\{ {\left| {m,n, + , - } \right\rangle ,\left| {m,n, -,  + } \right\rangle } \right\}$ and the eigenstate is
\begin{equation}
{\left| \psi  \right\rangle _{m, n}} = \frac{1}{{\sqrt 2 }}(\left| {m, n, + , - } \right\rangle  - \left| {m, n, -,  + } \right\rangle ),
\end{equation}
which is the famous ¡°dark state¡± \cite{Rodriguez-Lara2014,Peng2015}, where the spin singlet is decoupled from the photon field.

Now, take $E = {2\omega _1} + 2\omega $ for an example, the subspace is a six-dimension vector after the truncation in Eq.(\ref{truncation}),
\[\left\{ {\left| {0,0, + , - } \right\rangle ,\left| {0,0, -  + } \right\rangle ,\left| {1,0, - , - } \right\rangle ,\left| {0,1, + , + } \right\rangle ,\left| {1,1, + , - } \right\rangle ,\left| {1,1, - , + } \right\rangle } \right\}.\]
The Hamiltonian  ${H_{2{\omega _1} + 2\omega }}$ is written as
\begin{eqnarray}
{H_{2{\omega _1} + 2\omega }} = \left( {\begin{array}{*{20}{c}}
0&0&{{g_1}^\prime }&{{K_2}}&0&0\\
0&0&{{g_2}^\prime }&{{K_1}}&0&0\\
{{g_1}^\prime }&{{g_2}^\prime }&{{\omega _1} + \omega }&0&{{K_1}}&{{K_2}}\\
{{K_2}}&{{K_1}}&0&{{\omega _1} + \omega }&{{g_2}^\prime }&{{g_1}^\prime }\\
0&0&{{K_1}}&{{g_2}^\prime }&{2{\omega _1} + 2\omega }&0\\
0&0&{{K_2}}&{{g_1}^\prime }&0&{2{\omega _1} + 2\omega }
\end{array}} \right).
\end{eqnarray}
It is straightforward to calculate the eigenstate
\begin{eqnarray}
\left| \psi  \right\rangle _{2{\omega _1} + 2\omega } =&& \frac{{g'}}{K}\left( {\frac{{ - {\omega _1} - \omega }}{{g'}} - \frac{{g'}}{{ - 2{\omega _1} - 2\omega }} - \frac{K}{{ - 2{\omega _1} - 2\omega }} - \frac{{{K^2}}}{{g'}}\frac{{\left( { - {\omega _1} - \omega } \right)}}{{\left( {{{g'}^2} - {K^2}} \right)}}} \right)\left| {0,0, + , - } \right\rangle \nonumber\\
 &&+ \frac{{g'}}{K}\left( {\frac{{ - {\omega _1} - \omega }}{{g'}} - \frac{{g'}}{{ - 2{\omega _1} - 2\omega }} - \frac{K}{{ - 2{\omega _1} - 2\omega }} - \frac{{{K^2}}}{{g'}}\frac{{\left( { - {\omega _1} - \omega } \right)}}{{\left( {{{g'}^2} - {K^2}} \right)}}} \right)\left| {0,0, -  + } \right\rangle \nonumber\\
 && - \frac{{g'}}{K}\left| {1,0, - , - } \right\rangle  + \left| {0,1, + , + } \right\rangle  + \left| {1,1, + , - } \right\rangle  - \left| {1,1, - , + } \right\rangle.
\end{eqnarray}
Following the truncation of Eq.(\ref{truncation}), for $E = {\omega _1}m + \omega n$, the correspondence Hamiltonian $H_{\omega _1 m + \omega n}$ is a $[2(m+n-1)\times 2(m+n-1)]$ matrix in quasi-exact eigenstates. So, it is interesting, the quasi-exact solution of the system can be clearly found from the algebraic structure of Hamiltonian.
\section{CONCLUSION}
In summary, quantum Rabi model has a simple form of expression, but it is difficult to be solved, and for the two-qubit Rabi model, the analytical solution is more difficult. We have studied two exceptional solution of two-qubit Rabi model in two special situations. Firstly, if the frequencies of two qubit$\Delta_1, \Delta_1$ and photon field $\omega$ satisfy resonant relation, the two-qubit Rabi model can be mapped into the solvable formation in qubit-photon¡¯s Fock states space. If the resonance condition can be realized in the experiment, then it will be a method to test Rabi model. Secondly, we have studied the two-qubit quantum Rabi model in reservoir. Here, two-qubit system consist of two qubits and photon field, if the system interact with environments, the Boson field of reservoir, coupling between qubits and reservoir, and coupling between reservoir field and photon field should be considered additionally. In single Lorentzian model, the interaction between the reservoir and the quantum system in the whole spectrum is represented by the interaction of a pseudomode at the singular point according to the pseudomode approach, the Hamiltonian describing the interaction between two qubits and the reservoir and the optical field is a complicated matrix in complete Fock states space. Then, the algebraic structure of Hamiltonian is analyzed in the photon number space, a closed quasi-exact eigenstates space is found, and the quasi-exact solution can be clearly found from the algebraic structure of Hamiltonian. Our results as well as any finding basis on them provide a program to test the Rabi model.

\begin{acknowledgments}
This work was supported partly by Natural Science Foundation of Hebei Province,
Research Project of Universities in Hebei Province,
Grant Number Z2012027, and the Fundamental Research Funds for the Central
Universities.
\end{acknowledgments}

\appendix
\section{The effective Hamiltonian}
In the eigenstates space of ${\sigma _{1x}}$ and ${\sigma _{2x}}$, ${\sigma _{1x}}{\left|  \pm  \right\rangle _1}{\rm{ = }} \pm {\left|  \pm  \right\rangle _1},{\sigma _{2x}}{\left|  \pm  \right\rangle _2}{\rm{ = }} \pm {\left|  \pm  \right\rangle _2}$, defining
\[{\tau _{1z}} = {\left|  +  \right\rangle _1}_1\left\langle {\rm{ + }} \right| - {\left|  -  \right\rangle _1} _1\left\langle  -  \right|, {\tau _{2z}} = {\left|  +  \right\rangle _2} _{\rm{2}}\left\langle {\rm{ + }} \right| - {\left|  -  \right\rangle _2}_2\left\langle  -  \right|,\]
\[{\tau _{1 + }} = {\left|  +  \right\rangle _{1}}_1\left\langle  -  \right|, {\tau _{2 + }}={\left|  +  \right\rangle _{2}}_2\left\langle  -  \right|, {\kern 1pt} {\tau _{1 - }} = {\left|  -  \right\rangle _{1}} _1\left\langle  +  \right|, {\kern 1pt} {\tau _{2 - }} = {\left|  -  \right\rangle _{2}} _2\left\langle  +  \right|,\]
The Pauli spin operators is
\[\begin{array}{l}
{\sigma _{1z}} =  - ({\tau _{1 + }} + {\tau _{1 - }}),{\sigma _{1y}} =  - i({\tau _{1 + }} - {\tau _{1 - }}),{\sigma _{1x}} \to {\tau _{1z}}\\
{\sigma _{2z}} =  - ({\tau _{2 + }} + {\tau _{2 - }}),{\sigma _{2y}} =  - i({\tau _{2 + }} - {\tau _{2 - }}),{\sigma _{2x}} \to {\tau _{2z}}
\end{array}\]

In Eq.(4), $consh[2\lambda ({a^ {\dag} } + a)]$ and $sinh[2\lambda ({a^ {\dag} } + a)]$ are the even and odd functions respectively, the terms  and  can be expanded as
\begin{subequations}
\begin{eqnarray}
consh[2\lambda ({a^ {\dag} } + a)] &&= {G_0}(N) + {G_1}(N){a^ {\dag} }^2 + {a^2}{G_1}(N) +  \cdots, \\
sinh[2\lambda ({a^ {\dag} } + a)] &&={{\rm{F}}_1}(N){a^ {\dag} } - a{F_1}(N) + {F_2}(N)a +  \cdots.
\end{eqnarray}
\end{subequations}
in which, ${G_i}(N)$ and ${{\rm{F}}_{\rm{j}}}(N),i = 0,1,2\cdots $, with $N = {a^\dag }a$ are the coefficients that depend on the dimensionless parameter $\lambda$ and the photon number $n$. In Fock space of Eq.{7a}, the Hamiltonian matrix takes the formation as
\begin{equation}
_ + \left\langle j \right|{H_E}{\left| j \right\rangle _ + } = \left[ {\begin{array}{*{20}{c}}
 \ddots & \vdots & \vdots & \vdots & \vdots & \vdots & \vdots & {\mathinner{\mkern2mu\raise1pt\hbox{.}\mkern2mu
 \raise4pt\hbox{.}\mkern2mu\raise7pt\hbox{.}\mkern1mu}} \\
 \cdots &{{H_{11}}}&{{H_{12}}}&{{H_{13}}}&{{H_{14}}}&{{H_{15}}}&{{H_{16}}}& \cdots \\
 \cdots &{{H_{21}}}&{{H_{22}}}&{{H_{23}}}&{{H_{24}}}&{{H_{25}}}&{{H_{26}}}& \cdots \\
 \cdots &{{H_{31}}}&{{H_{32}}}&{{H_{33}}}&{{H_{34}}}&{{H_{35}}}&{{H_{36}}}& \cdots \\
 \cdots &{{H_{41}}}&{{H_{42}}}&{{H_{43}}}&{{H_{44}}}&{{H_{45}}}&{{H_{46}}}& \cdots \\
 \cdots &{{H_{51}}}&{{H_{52}}}&{{H_{53}}}&{{H_{54}}}&{{H_{55}}}&{{H_{56}}}& \cdots \\
 \cdots &{{H_{61}}}&{{H_{62}}}&{{H_{63}}}&{{H_{64}}}&{{H_{65}}}&{{H_{66}}}& \cdots \\
 {\mathinner{\mkern2mu\raise1pt\hbox{.}\mkern2mu
 \raise4pt\hbox{.}\mkern2mu\raise7pt\hbox{.}\mkern1mu}} & \vdots & \vdots & \vdots & \vdots & \vdots & \vdots & \ddots
\end{array}} \right]
\end{equation}
in which
\begin{eqnarray*}
{H_{11}} &&= \left\langle {2n+1, -, +} \right|{H_E}\left| {2n+1, -, +} \right\rangle  \\
&&= \omega (2n + 1) + \lambda _1^2\omega  + \lambda _{^2}^2\omega  + 2{\lambda _1}{g_1} + 2{\lambda _2}{g_2} - {\Delta _1}{G_0}(2n + 1) + {\Delta _2}{G_0}(2n + 1).\\
{H_{12}} &&= \left\langle {2n+1, -, +} \right|{H_E}\left| {2n,+1, +, -} \right\rangle  = {H_{21}} = \left\langle {2n+1, +, - } \right|{H_E}\left| {2n+1, -, +} \right\rangle \\
&& = 2{\lambda _2}{g_1} + 2{g_2}{\lambda _1} + 2{\lambda _1}{\lambda _2}\omega.\\
{H_{13}} &&= \left\langle {2n + 1, - , + } \right|{H_E}\left| {2n + 2, + , + } \right\rangle  = {H_{31}} = \left\langle {2n + 2, + , + } \right|{H_E}\left| {2n + 1, - , + } \right\rangle \\
&& = ({g_1} + {\lambda _1}\omega )\sqrt {2n + 2}  + {\Delta _1}{F_1}(2n + 2,2n + 1).\\
{H_{14}} &&= \left\langle {2n + 1, - , + } \right|{H_E}\left| {2n + 2, - , - } \right\rangle  = {H_{41 = }}\left\langle {2n + 2, - , - } \right|{H_E}\left| {2n + 1, - , + } \right\rangle \\
&& = ({g_2} + {\lambda _2}\omega )\sqrt {2n + 2}  - {\Delta _2}{F_1}^\prime (2n + 2,2n + 1).
\end{eqnarray*}
\begin{eqnarray*}
{H_{22}} &&= \left\langle {2n + 1, + , - } \right|{H_E}\left| {2n + 1, + , - } \right\rangle \\
 &&= (2n + 1)\omega  + \lambda _1^2\omega  + \lambda _{^2}^2\omega  + 2{\lambda _1}{g_1} + 2{\lambda _2}{g_2} + {\Delta _1}{G_0}(2n + 1) - {\Delta _2}{G_0}^\prime (2n + 1).\\
{H_{23}} &&= \left\langle {2n + 1, + , - } \right|{H_E}\left| {2n + 2, + , + } \right\rangle  = {H_{32}} = \left\langle {2n + 2, + , + } \right|{H_E}\left| {2n + 1, + , - } \right\rangle  \\
&&= ({g_2} + {\lambda _2}\omega )\sqrt {2n + 2}  + {\Delta _2}{F_1}^\prime (2n + 2,2n + 1).\\
{H_{24}} &&= \left\langle {2n + 1, + , - } \right|{H_E}\left| {2n + 2, - , - } \right\rangle  = {H_{42}} = \left\langle {2n + 2, - , - } \right|{H_E}\left| {2n + 1, + , - } \right\rangle \\
&& = ({g_1} + {\lambda _1}\omega )\sqrt {2n + 2}  - {\Delta _1}{F_1}(2n + 2,2n + 1).
\end{eqnarray*}
\begin{eqnarray*}
{H_{33}} &&= \left\langle {2n + 2, + , + } \right|{H_E}\left| {2n + 2, + , + } \right\rangle  \\
&&= (2n + 2)\omega  + \lambda _1^2\omega  + \lambda _{^2}^2\omega  + 2{\lambda _1}{g_1} + 2{\lambda _2}{g_2} + {\Delta _1}{G_0}(2n + 2) + {\Delta _2}{G_0}^\prime (2n + 2).\\
{H_{34}} &&= \left\langle {2n + 2, + , + } \right|{H_E}\left| {2n + 2, - , - } \right\rangle  = {H_{43}} = \left\langle {2n + 2, - , - } \right|{H_E}\left| {2n + 2, + , + } \right\rangle \\
&& = 2{\lambda _2}{g_1} + 2{g_2}{\lambda _1} + 2{\lambda _1}{\lambda _2}\omega. \\
{H_{35}} &&= \left\langle {2n + 2, + , + } \right|{H_E}\left| {2n + 3, - , + } \right\rangle  = {H_{53}} = \left\langle {2n + 3, - , + } \right|{H_E}\left| {2n + 2, + , + } \right\rangle \\
&& = ({g_1} + {\lambda _1}\omega )\sqrt {2n + 3}  - {\Delta _1}{F_1}(2n + 3,2n + 2).\\
{H_{36}} &&= \left\langle {2n + 2, + , + } \right|{H_E}\left| {2n + 3, + , - } \right\rangle  = {H_{63}} = \left\langle {2n + 3, + , - } \right|{H_E}\left| {2n + 2, + , + } \right\rangle \\
&& = ({g_2} + {\lambda _2}\omega )\sqrt {2n + 3}  - {\Delta _2}{F_1}^\prime (2n + 3,2n + 2).
\end{eqnarray*}
\begin{eqnarray*}
{H_{44}} &&= \left\langle {2n + 2, - , - } \right|{H_E}\left| {2n + 2, - , - } \right\rangle  \\
&&= (2n + 2)\omega  + \lambda _1^2\omega  + \lambda _{^2}^2\omega  + 2{\lambda _1}{g_1} + 2{\lambda _2}{g_2} - {\Delta _1}{G_0}(2n + 2) - {\Delta _2}{G_0}^\prime (2n + 2).\\
{H_{45}} &&= \left\langle {2n + 2, - , - } \right|{H_E}\left| {2n + 3, - , + } \right\rangle  = {H_{54}} = \left\langle {2n + 3, - , + } \right|{H_E}\left| {2n + 2, - , - } \right\rangle  \\
&&= ({g_2} + {\lambda _2}\omega )\sqrt {2n + 3}  + {\Delta _2}{F_1}^\prime (2n + 3,2n + 2).\\
{H_{46}} &&= \left\langle {2n + 2, - , - } \right|{H_E}\left| {2n + 3, + , - } \right\rangle  = {H_{64}} = \left\langle {2n + 3, + , - } \right|{H_E}\left| {2n + 2, - , - } \right\rangle \\
&& = ({g_1} + {\lambda _1}\omega )\sqrt {2n + 3}  + {\Delta _1}{F_1}(2n + 3,2n + 2).
\end{eqnarray*}
\begin{eqnarray*}
{H_{55}} &&= \left\langle {2n + 3, - , + } \right|{H_E}\left| {2n + 3, - , + } \right\rangle \\
 &&= \omega (2n + 3) + \lambda _1^2\omega  + \lambda _{^2}^2\omega  + 2{\lambda _1}{g_1} + 2{\lambda _2}{g_2} - {\Delta _1}{G_0}(2n + 3) + {\Delta _2}{G_0}(2n + 3).\\
{H_{56}} &&= \left\langle {2n + 3, - , + } \right|{H_E}\left| {2n + 3, + , - } \right\rangle  = {H_{65}} = \left\langle {2n + 3, + , - } \right|{H_E}\left| {2n + 3, - , + } \right\rangle  \\
&&= 2{\lambda _2}{g_1} + 2{g_2}{\lambda _1} + 2{\lambda _1}{\lambda _2}\omega. \\
{H_{66}} &&= \left\langle {2n + 3, + , - } \right|{H_E}\left| {2n + 3, + , - } \right\rangle  \\
&&= \omega (2n + 3) + \lambda _1^2\omega  + \lambda _{^2}^2\omega  + 2{\lambda _1}{g_1} + 2{\lambda _2}{g_2} + {\Delta _1}{G_0}(2n + 3) - {\Delta _2}{G_0}^\prime (2n + 3).
\end{eqnarray*}
\begin{eqnarray*}
{H_{15}} &&= \left\langle {2n + 1, - , + } \right|{H_E}\left| {2n + 3, - , + } \right\rangle  = {H_{51}} = \left\langle {2n + 3, - , + } \right|{H_E}\left| {2n + 1, - , + } \right\rangle  = 0\\
{H_{16}} &&= \left\langle {2n + 1, - , + } \right|{H_E}\left| {2n + 3, + , - } \right\rangle  = {H_{61}} = \left\langle {2n + 3, + , - } \right|{H_E}\left| {2n + 1, - , + } \right\rangle  = 0\\
{H_{25}} &&= \left\langle {2n + 1, + , - } \right|{H_E}\left| {2n + 3, - , + } \right\rangle  = {H_{52}} = \left\langle {2n + 3, - , + } \right|{H_E}\left| {2n + 1, + , - } \right\rangle  = 0\\
{H_{26}} &&= \left\langle {2n + 1, + , - } \right|{H_E}\left| {2n + 3, + , - } \right\rangle  = {H_{62}} = \left\langle {2n + 3, + , - } \right|{H_E}\left| {2n + 1, + , - } \right\rangle  = 0
\end{eqnarray*}
It is straightforward to calculate that
\begin{subequations}
\begin{eqnarray}
&&G_{_0}^{\left( 1 \right)}(n){\rm{ = }}\left\langle n \right|consh[2{\lambda _1}({a^ {\dag} } - a)]\left| n \right\rangle  = {e^{ - 2\lambda _1^2}}{L_n}(4\lambda _1^2)\\
&&G_{_0}^{\left( 2 \right)}(n){\rm{ = }}\left\langle n \right|consh[2{\lambda _2}({a^ {\dag} } - a)]\left| n \right\rangle  = {e^{ - 2\lambda _{^2}^2}}{L_n}(4\lambda _{^2}^2)\\
&&F_{_1}^{\left( 1 \right)}(n + 1,n) = \left\langle {n + 1} \right|\sin h[2{\lambda _1}({a^ {\dag} } - a)]\left| n \right\rangle {\rm{ = }}\frac{{2{\lambda _1}{e^{ - 2\lambda _1^2}}L_n^1(4\lambda _1^2)}}{{\sqrt {n + 1} }}\\
&&F_{_1}^{\left( 2 \right)}(n + 1,n) = \left\langle {n + 1} \right|\sin h[2{\lambda _2}({a^ {\dag} } - a)]\left| n \right\rangle {\rm{ = }}\frac{{2{\lambda _2}{e^{ - 2\lambda _{^2}^2}}L_n^1(4\lambda _{^2}^2)}}{{\sqrt {n + 1} }}
\end{eqnarray}
\end{subequations}
when $\lambda$ is small, the Laguerre polynomial is given by
\[L_n^1(4{\lambda ^2}) \approx n + 1,{L_n}(4{\lambda ^2}) \approx 1.\]
If the dimensionless parameter $\lambda_1$ and $\lambda_2$ are chosen as Eq.(8), then $H_{13}$ and $H_{14}$ are zero. And the resonant relation Eq.(9) makes
\begin{equation}
2{\lambda _2}{g_1} + 2{g_2}{\lambda _1} + 2{\lambda _1}{\lambda _2}\omega  = 0.
\end{equation}
Then, the Hamiltonian takes the formation of a block diagonal matrix as Eq.(\ref{block diagonal Matrix}).

\bibliography{twoqubitRabi20160907}

\end{document}